\documentclass[10pt]{article}
\usepackage{epsfig,dsfont}

\begin{document}
\sffamily

\thispagestyle{empty}
\vspace*{15mm}

\begin{center}

{\LARGE 
Flux representation of an effective \vskip3mm
Polyakov loop model for QCD thermodynamics}
\vskip15mm
Christof Gattringer 
\vskip8mm
Institut f\"ur Physik, Karl-Franzens-Unversit\"at Graz \\
Universit\"atsplatz 5, 8010 Graz, Austria 
\vskip5mm
 
\end{center}
\vskip30mm

\begin{abstract}
We discuss an effective Polyakov loop model for QCD thermodynamics with a chemical
potential. Using high temperature expansion techniques the partition sum is mapped
exactly onto the partition sum of a flux model. In the flux representation the
complex action problem is resolved and a simulation with worm-type algorithms
becomes possible also at finite chemical potential. 
\end{abstract}

\vskip35mm
\begin{center}
{\it To appear in Nuclear Physics B.}
\end{center}

\setcounter{page}0
\newpage
\noindent
{\Large Introductory remarks}
\vskip3mm
\noindent
With the running and upcoming experiments at BNL, CERN, GSI and Dubna the amount
of experimental facts about the QCD phase diagram will increase considerably in the
near future. Also
the theoretical side is challenged to contribute to our understanding of QCD
thermodynamics, a task which is rather demanding due to the non-perturbative
nature of the problem. In principle lattice QCD is a suitable non-perturbative
approach, but at finite chemical potential the complex phase problem
considerably limits the applicability of Monte Carlo methods for a numerical
evaluation of the path integral.  

A tempting idea to overcome the complex phase
problem is to search for a transformation to new degrees of freedom where the
partition function  becomes a sum over configurations with real and positive
weights. Such representations with real and positive weights 
can then be used in a Monte Carlo calculation. Several
examples for this kind of transformations in various models can be found in the literature, but for
QCD we are still far from identifying suitable degrees of freedom.

In this article we contribute to the enterprise of developing new and real
representations for QCD related systems with a quark chemical potential $\mu$ by considering an
effective theory for Polyakov loop variables. The Polyakov loop has
the interpretation of a static color source and in pure gauge theory its 
expectation value serves as an order parameter for the deconfinement transition.
More specifically it is an order parameter for center symmetry which is broken
spontaneously at the deconfinement transition of pure gauge theory.
Although in full QCD the underlying center symmetry is broken explicitly by the
quarks, also there the Polyakov loop may be used to determine the crossover to
deconfinement.  

The idea of using effective theories for the Polyakov loop to
describe the transitions in pure gluodynamics and in QCD goes back to the late seventies, and since then,
both the understanding of sub-leading terms in the effective action, as well as the simulation 
techniques have been improved considerably \cite{models1}--\cite{effmodelCL}.

The effective theory for the Polyakov loop considered here may be derived from QCD
using strong coupling expansion for the gauge action and hopping expansion for the
fermion determinant. The leading center symmetric and center symmetry breaking
terms are taken into account, as well as the chemical potential $\mu$. For
$\mu \neq 0$ the action becomes complex, i.e., in its standard representation the
theory inherits the complex phase problem from QCD. 

In this work we apply high
temperature expansion techniques and derive a flux representation which is free of
the complex phase problem. The new degrees of freedom are integer valued flux
variables attached to the links of the lattice and integer valued monomer
variables at the sites. The flux-monomer configurations are subject to constraints
which enforce the total flux at each site to be a multiple of three. The weight
factors for the admissible configurations are given in closed form and turn out to
be real and positive. 
Thus, in the flux representation a Monte Carlo simulation of 
the system with generalized Prokof'ev-Svistunov worm 
algorithms \cite{worm} becomes possible and should allow for an effective 
numerical treatment without any complex phase problems.
Finally, we show how observables can be expressed in terms of the
flux and monomer variables. 

\newpage
\noindent 
{\Large The effective theory and its high temperature expansion}
\vskip3mm
\noindent
The effective Polyakov loop theory we consider is described by the action
\begin{eqnarray}
S[P] & \; = \; & - \, \tau \sum_{x} \sum_{\nu=1}^3 
\Big[ \, \mathrm{Tr}\, P(x) \, \mathrm{Tr} \, P(x\!+\!\hat{\nu})^\dagger \, + \, 
\mathrm{Tr} \, P(x)^\dagger \, \mathrm{Tr} \, P(x\!+\!\hat{\nu}) \, \Big] 
\nonumber \\
& & - \, 
\sum_{x} \Big[ \, \eta \, \mathrm{Tr}\, P(x) \, + \, \overline{\eta} \; \mathrm{Tr}\,
P(x)^\dagger \, \Big]\; .
\label{action_original}
\end{eqnarray}
In this standard representation the degrees of freedom are the SU(3) valued
Polyakov variables $P(x)$ attached to the sites $x$ of a three-dimensional cubic
lattice which we consider to be finite with periodic boundary conditions. 
By $\hat{\nu}$ we denote the unit vector in $\nu$-direction, with $\nu = 1,2,3$.
The first term of the action, which may be obtained from a strong
coupling expansion, is a nearest neighbor interaction of the
traced Polyakov loops. This expansion also establishes $\tau$ to be an increasing
function of the QCD temperature, and for simplicity we refer to $\tau$ as
{\it temperature}. The term in the second line of (\ref{action_original}) can be
derived from the fermion determinant using hopping expansion. The parameters 
$\eta$ and $\overline{\eta}$ are related to the 
amplitude $\kappa$ and the chemical potential $\mu$ via
\begin{equation}
\eta \; = \; \kappa \, e^{\, \mu} \quad , \qquad \overline{\eta} \; = \; 
\kappa \, e^{-\mu} \; .
\label{etaetabar}
\end{equation}
The hopping expansion identifies 
$\kappa$ to be proportional to a power of the hopping parameter and thus $\kappa$ is a 
function of the quark mass $m$ which decreases with increasing
$m$. For $N_f$ flavors of mass degenerate quarks $\kappa$ is simply proportional
to $N_f$. The terms $e^{\mu} \, \mathrm{Tr}\,P(x), \, e^{-\mu}\, 
\mathrm{Tr}\, P(x)^\dagger$ in (\ref{action_original}), (\ref{etaetabar}) are the
leading $\mu$-dependent terms of the fermion determinant\footnote{Actually in the
fermion determinant the chemical potential appears rescaled with the temporal
extent $N_t$ of the lattice, but in order to simplify the notation used for the
effective theory we omit this
factor.}.

From Eq.~(\ref{action_original}) one immediately sees that the effective theory has a complex
phase problem: For $\mu \neq 0$ one has $\eta \neq \overline{\eta}$, which leads to a
non-vanishing imaginary contribution to the action given by
$i \sum_x (\overline{\eta} - \eta) \, \mathrm{Im} \, P(x)$. Thus the Boltzmann factor 
$\exp(-S[P])$ obtains a complex phase and cannot be used as a probability weight in a 
Monte Carlo calculation.  

Effective theories of the type (\ref{action_original}) were studied before in 
various contexts. For the case of vanishing $\kappa$ an important 
line of research is the determination of sub-leading terms in the action to get
the effective theory ready for quantitative analysis -- see, e.g.,
\cite{models5}--\cite{models6}. For the case of non-zero chemical potential 
an important contribution is \cite{effmodelCL}, where the theory described by 
(\ref{action_original}) was studied with complex Langevin techniques. A future
numerical simulation with the flux representation presented here should be able
to check the results \cite{effmodelCL} and shed light on the reliability of the 
complex Langevin method.

A reduced version of the model (\ref{action_original}), where the traced Polyakov loops are replaced by
center elements, was studied as well \cite{models2}--\cite{DelEveGat}. In this
simpler
case a flux representation free of complex phases has been known for a long time
\cite{models2} and various numerical simulations with different techniques were
presented \cite{models2}--\cite{DelEveGat}.

The grand canonical partition function of the model described by 
(\ref{action_original}) is obtained by integrating the Boltzmann
factor $e^{-S[P]}$ over all configurations of the Polyakov loop variables. The
corresponding measure is a product over the reduced Haar measures $d P(x)$ at the sites
$x$. Thus
\begin{equation}
Z \; = \; \int \! \prod_x d P(x) \; e^{-S[P]} \; = \;  \int \! D[P] \; e^{-S[P]} \; ,
\end{equation}
where in the second step we have introduced the shorthand notation 
 $D[P] = \prod_x d P(x) $ for the product measure.

The first steps towards the flux representation are writing the Boltzmann factors
in a factorized form and an expansion of the remaining exponentials. This step
corresponds to an expansion in $\tau$. In spin system language $\tau$ would have
to be identified with the inverse temperature $\beta$ and thus our expansion is
equivalent to high temperature expansion in statistical mechanics. We obtain the
following form of the partition sum:
\begin{eqnarray}
Z\!\! &\!\! =\!\!\! & \!\!\int\!\! D[P]\! \left( \prod_{x,\nu}  
e^{\tau \, \mathrm{Tr}\, P(x) \, \mathrm{Tr} \, P(x + \hat{\nu})^\dagger} \,
e^{\tau \, \mathrm{Tr}\, P(x)^\dagger \, \mathrm{Tr} \, P(x + \hat{\nu})} \!\right)
\!\!\! \left( \prod_x 
e^{\eta \, \mathrm{Tr}\, P(x)} \, 
e^{\overline{\eta} \, \mathrm{Tr}\, P(x)^\dagger} \!\right)
\nonumber \\
&\!\! =\!\!\! & \!\!\int\!\! D[P]\! \left( \prod_{x,\nu} \left[\sum_{l_{x,\nu} =
0}^\infty \frac{\tau^{l_{x,\nu}}}{l_{x,\nu}!} 
\Big(\mathrm{Tr}\, P(x)\Big)^{l_{x,\nu}} \Big(\mathrm{Tr} \, P(x +
\hat{\nu})^\dagger\Big)^{l_{x,\nu}} \right] \right)
\nonumber \\
&& \hspace{6mm} \times \left( \prod_{x,\nu} \left[\sum_{\overline{l}_{x,\nu} =
0}^\infty \frac{\tau^{\overline{l}_{x,\nu}}}{\overline{l}_{x,\nu}!} 
\Big(\mathrm{Tr}\, P(x)^\dagger\Big)^{\overline{l}_{x,\nu}} \Big(\mathrm{Tr} \, P(x +
\hat{\nu})\Big)^{\overline{l}_{x,\nu}} \right] \right)
\nonumber \\
&& \times \Bigg( \prod_{x} \Bigg[ \sum_{s_x =
0}^\infty \frac{\eta^{s_x}}{s_x!}
\Big(\mathrm{Tr}\, P(x)\Big)^{s_x}\Bigg] \Bigg)
\Bigg( \prod_{x} \Bigg[ \sum_{\overline{s}_x =
0}^\infty \frac{\overline{\eta}^{\overline{s}_x}}{\overline{s}_x!}
\Big(\mathrm{Tr}\, P(x)^\dagger\Big)^{\overline{s}_x}\Bigg] \Bigg) \; .
\label{step1}
\end{eqnarray}
In the end the partition function will be a sum over 
configurations of the expansion coefficients $l_{x,\nu}, \overline{l}_{x, \nu}$ 
for the nearest neighbor terms,
and the expansion coefficients $s_{x}, \overline{s}_{x}$ for the site terms.
We will refer to $l_{x,\nu}, \overline{l}_{x, \nu}$ as {\it flux variables} and to 
$s_{x}, \overline{s}_{x}$ as {\it monomer variables}. For the sum over flux- and
monomer configurations we introduce the shorthand notation
\begin{equation}
\sum_{\{l,\overline{l},s,\overline{s}\}} \; = \; 
\left(\prod_{x,\nu} \; \sum_{l_{x,\nu} =
0}^\infty \; \sum_{\overline{l}_{x,\nu} =
0}^\infty \, \right) \left( \prod_{x} \sum_{s_x =
0}^\infty \; \sum_{\overline{s}_x =
0}^\infty \, \right) \;.
\end{equation}
Using this notation, rearranging the products in (\ref{step1}), and writing
the integral over all configurations of the Polyakov loops $P(x)$ again in its
factorized form, the
partition sum reads
\begin{eqnarray}
Z \!& \!\! = \!\!& \!\!\! \sum_{\{l,\overline{l},s,\overline{s}\}} \! \left(
\prod_{x,\nu} \; 
\frac{\tau^{l_{x,\nu} + \overline{l}_{x,\nu}}}{l_{x,\nu}! \; \overline{l}_{x,\nu}!}
\right) \left( \prod_{x} \;
\frac{\eta^{s_x} \; \overline{\eta}^{\overline{s}_x}}{s_x! \; \overline{s}_x!}  
\right)
\\
&&\! \! \times \! \left(\! \prod_x \int \!\! d P(x) \, \Big(\mathrm{Tr} \, P(x)\Big)^{ \sum_\nu [
l_{x,\nu} + \overline{l}_{x-\hat{\nu},\nu} ] + s_x}  
\Big(\mathrm{Tr} \, P(x)^\dagger\Big)^{ \sum_\nu [
\overline{l}_{x,\nu} + l_{x-\hat{\nu},\nu} ] + \overline{s}_x} \right)
\nonumber \\
& \!\! = \!\!& \!\!\! \sum_{\{l,\overline{l},s,\overline{s}\}} \!\!
W[l,\overline{l},s,\overline{s}] \; \prod_x \, {I} \bigg( \sum_\nu [
l_{x,\nu} + \overline{l}_{x-\hat{\nu},\nu} ] + s_x \bigg| \sum_\nu [
\overline{l}_{x,\nu} + l_{x-\hat{\nu},\nu} ] + \overline{s}_x \bigg) \; .
\nonumber
\end{eqnarray}
In the last step we introduced the weight factor for a configuration of flux- and
monomer variables,
\begin{equation}
W[l,\overline{l},s,\overline{s}] \; = \; \left( \prod_{x,\nu} \;
\frac{\tau^{l_{x,\nu} + \overline{l}_{x,\nu}}}{l_{x,\nu}! \; \overline{l}_{x,\nu}!}
\right) \left( \prod_{x} \;
\frac{\eta^{s_x} \; \overline{\eta}^{\overline{s}_x}}{s_x! \; \overline{s}_x!}  
\right)
\; ,
\end{equation}
and for the remaining SU(3) integrals at the sites use the abbreviation
\begin{equation}
{I} (n | m) \; = \; \int dP \, \Big( \mathrm{Tr} \, P \Big)^n \, 
\Big( \mathrm{Tr} \, P^\dagger \Big)^m \; .                               
\end{equation}
Here $n$ and $m$ are non-negative integers and $\int\! dP$ denotes the integration
over SU(3) Haar measure.

\vskip8mm
\noindent
{\Large Solving the SU(3) integrals}
\vskip3mm
\noindent 
In this section we evaluate the remaining SU(3) integrals ${I} (n | m)$. The
corresponding generating function is the integral
\begin{equation}
{G} (u | v) \; = \; \int dP \; e^{\,u \, \mathrm{Tr} \, P } \;
e^{\,v \, \mathrm{Tr} \, P^\dagger} \; = \; 
\sum_{n,m = 0}^\infty \frac{u^n}{n!} \, \frac{v^m}{m!} \; {I} (n | m) \; ,
\label{haarint1}
\end{equation}
in other words the $I(n|m)$ are the moments of the one link integral
${G} (u | v)$. We start our derivation of the ${I} (n | m)$
with an expression for 
${G} (u | v)$ given in \cite{wipfsu3},
$$
{G} (u | v) \; = \; \sum_{p,q = 0}^\infty \frac{2}{(p\!+\!q\!+\!1)! \, 
(p\!+\!q\!+\!2)! \, q!} \left( 3(p\!+\!q\!+\!1) \atop p \right) \, (uv)^p \, (u^3 + v^3)^q \; .
$$ 
We use the binomial formula to evaluate $(u^3 + v^3)^q$ and organize the terms
with respect to the monomials $u^n \, v^m$ to obtain,
\begin{eqnarray}
{G} (u | v)\!\! & = & \!\!\sum_{p,q = 0}^\infty \frac{2}{(p\!+\!q\!+\!1)! \, 
(p\!+\!q\!+\!2)! \, q!} \left( 3(p\!+\!q\!+\!1) \atop p \right) \, \sum_{j=0}^q \left( q \atop j
\right) \, u^{p+3j}
v^{p+3q-3j} \nonumber \\
& = & \!\!\! \sum_{n,m=0}^\infty \!\!
\frac{u^n}{n!} \, \frac{v^m}{m!} \;\; \sum_{p,q = 0}^\infty \sum_{j=0}^q
\frac{ 2 n! m! \; \delta_{n,p+3j} \, \delta_{m,p+3q-3j}}{(p\!+\!q\!+\!1)! \, 
(p\!+\!q\!+\!2)! \, q!} \left( 3(p\!+\!q\!+\!1) \atop p \right) \! \left( q \atop j
\right).
\nonumber
\end{eqnarray}
Comparing this expression with the second form of the generating function 
(\ref{haarint1}), we identify
$$
{I} (n | m) = \! \sum_{p,q = 0}^\infty \sum_{j=0}^q
\delta_{n,p+3j} \, \delta_{m,p+3q-3j} \, \frac{ 2 n! m! \;}{(p\!+\!q\!+\!1)! \, 
(p\!+\!q\!+\!2)! \, q!} \left( 3(p\!+\!q\!+\!1) \atop p \right) \!\! \left( q \atop j
\right)\!.
$$
The two Kronecker deltas can now be used to reduce this expression to a finite
sum. The first one implies $p = n - 3j$, and since $p \geq 0$, we find
$n \geq 3j$ and obtain an upper bound for the sum over $j$ given by
$j \leq \lfloor n/3 \rfloor$, where by $\lfloor x \rfloor$ we denote the floor 
function\footnote{$\lfloor x \rfloor$ is the integer with $x - 1 < \lfloor x
\rfloor \leq x$.}. Thus after performing
the sum over $p$ we have
$$
{I} (n | m) \; = \; \sum_{q = 0}^\infty \; \sum_{j=0}^{\lfloor n/3 \rfloor}
\frac{ \delta_{m,n+3q-6j} \; 2 n! m! \;}{(n\!-\!3j\!+\!q\!+\!1)! \, 
(n\!-\!3j\!+\!q\!+\!2)! \, q!} \left( 3(n\!-3j\!+\!q\!+\!1) \atop n\!-\!3j \right) \!\! \left( q \atop j
\right)\!.
$$
The remaining Kronecker delta may be written as $\delta_{n-m,6j-3q}$, which makes
explicit an important property of the integrals ${I} (n | m)$, the triality
constraint:
\begin{equation}
{I} (n | m) \; \neq \; 0 \quad , \qquad \mbox{only if} \qquad (n-m) \, \mathrm{mod}
\, 3 \; = \; 0\; .
\label{triality}
\end{equation}
The Kronecker delta $\delta_{n-m,6j-3q}$ implies $q = 2j - (n-m)/3$. Using this
in the second binomial factor gives $\left( q \atop j \right) = 
\left( 2j - (n-m)/3 \atop j \right)$. In order to obtain a non-zero result
for this binomial, the upper argument may not be smaller than the lower one, i.e.,
$2j - (n-m)/3 \geq j$. This implies $j \geq (n-m)/3$, and since $j$ also must be
non-negative, we obtain the lower bound $j \geq \max\, (0,(n-m)/3)$. Performing 
the sum over $q$ and using the lower bound for $j$ we obtain the 
${I} (n | m)$ as finite sums
\begin{equation}
{I} (n | m) = 
  \sum_{j= \max\, (0,\frac{n\!-\!m}{3}) }^{\lfloor n/3 \rfloor} \!
\frac{ T(n-m) \;\; 2 \, n!\, m! \;\;
\left( 3(n-j-\frac{n-m}{3}+1) \atop n-3j \right) \!\! 
\left( 2j-\frac{n-m}{3} \atop j
\right) }{
(n\!-\!j -\frac{n-m}{3} + 1)! \, 
(n\!-\!j -\frac{n-m}{3} + 2)! \, 
(2j - \frac{n-m}{3})!} \; .
\label{inm_final}
\end{equation}
Here we have introduced the triality function
\begin{equation}
T(n) \; = \; \left\{ \begin{array}{cc}
1 & \mbox{for} \;\; n \, \mathrm{mod}\, 3 \; = \; 0 \; ,
\\
0 & \mbox{else} \; .
\end{array} \right.
\end{equation}
The final result (\ref{inm_final}) expresses the moments $I(n|m)$ as finite
sums. In \cite{wipfsu3} recursion relations that relate different $I(n|m)$
were presented and in an appendix the lowest moments for $n,m \leq 10$ 
are listed. We compared the results from our explicit expression 
(\ref{inm_final}) to these values from the recursion relation and found
agreement.

\vskip8mm
\noindent
{\Large Final form of the partition sum and graphical representation}
\vskip3mm
\noindent 
Using the results from the last section we obtain our final expression for the
partition sum in terms of flux and monomer variables,
\begin{equation}
Z \; = \sum_{\{l,\overline{l},s,\overline{s}\}} \!\!
{\cal W}[l,\overline{l},s,\overline{s}] \; {\cal C}[l,\overline{l},s,\overline{s}]
\; .
\label{zfinal}
\end{equation}
The first term under the sum is the total weight factor assigned to a configuration
of fluxes and monomers,
\begin{eqnarray}
{\cal W}[l,\overline{l},s,\overline{s}] & = &
\left(
\prod_{x,\nu} \; 
\frac{\tau^{l_{x,\nu} + \overline{l}_{x,\nu}}}{l_{x,\nu}! \; \overline{l}_{x,\nu}!}
\right) \left( \prod_{x} \;
\frac{\eta^{s_x} \; \overline{\eta}^{\overline{s}_x}}{s_x! \; \overline{s}_x!}  
\right) \, {\cal I}[l,\overline{l},s,\overline{s}] \qquad \mbox{with}
\nonumber \\
{\cal I}[l,\overline{l},s,\overline{s}] & = &  
\prod_x \, {I} \bigg( \sum_\nu [
l_{x,\nu} + \overline{l}_{x-\hat{\nu},\nu} ] + s_x \bigg| \sum_\nu [
\overline{l}_{x,\nu} + l_{x-\hat{\nu},\nu} ] + \overline{s}_x \bigg) \, . \qquad
\label{weightfinal}
\end{eqnarray}
The second factor under the sum in (\ref{zfinal}) is the constraint,
\begin{equation}
{\cal C}[l,\overline{l},s,\overline{s}] \; = \; 
\prod_x \, T \bigg( \sum_\nu \Big[
(l_{x,\nu} - \overline{l}_{x,\nu}) - 
(l_{x-\hat{\nu},\nu} - \overline{l}_{x-\hat{\nu},\nu}) \Big] + (s_x - \overline{s}_x) \bigg)
\; .
\label{constraintfinal}
\end{equation}
The constraint is a product over the sites $x$ and at each site the 
triality function $T$ enforces the combination
\begin{equation}
f_x \; = \; \sum_\nu \Big[
(l_{x,\nu} - \overline{l}_{x,\nu}) - 
(l_{x-\hat{\nu},\nu} - \overline{l}_{x-\hat{\nu},\nu}) \Big] + (s_x - \overline{s}_x)
\label{localflux}
\end{equation}
of flux and monomer variables to be a multiple of 3. The expression $f_x$ is the
total net flux at site $x$, including the contributions from the monomers.

It is instructive to compare the flux representation (\ref{zfinal}) -- (\ref{constraintfinal}) 
to the flux representation \cite{models1b,DelEveGat} for a simpler effective theory where the
Polyakov loop variables $\mathrm{Tr} P(x)$ are replaced by $\mathds{Z}_3$ center elements 
$P(x) \in \{1, e^{i2\pi/3}, e^{-i2\pi/3} \}$. The main difference is that in the 
discrete $\mathds{Z}_3$-case for the dimer and monomer variables only the three values $-1, 0 , +1$
are necessary. Here, where the dynamical variables are $\mathrm{Tr} P(x)$, i.e., continuous
degrees of freedom, integers with an infinite range are necessary for the monomers and dimers.
The triality constraint for the flux $f_x$ at a site $x$ is essentially the same in the
two models, with the difference, that
for the $\mathds{Z}_3$ case again only a finite number of values is possible, $f_x \in
\{-6,-3,0,3,6\}$. Finally, for the effective theory studied here, the weight 
(\ref{weightfinal}) contains the factors $I(n|m)$ which reflect the non-abelian nature of the
degrees of freedom $\mathrm{Tr} P(x)$.

\begin{figure}[h]
\vskip5mm
\begin{center}
\includegraphics[height=27mm,clip]{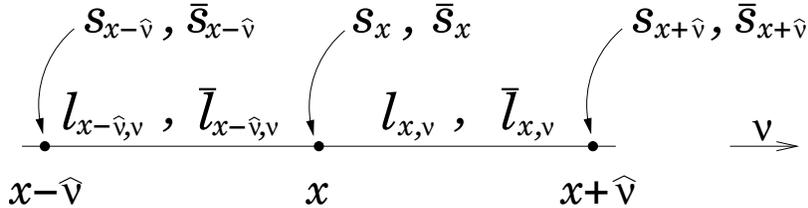}
\end{center}
\caption{Assignment of the flux variables $l_{x,\nu}, \overline{l}_{x,\nu}$, and
the monomer variables $s_{x}, \overline{s}_{x}$ to the links and sites of the
lattice. For clarity we display only one of the three possible directions.   
\label{variablesgrafical}}
\end{figure}

We now present a graphical representation for the flux and monomer
variables an discuss the triality constraint for the net flux $f_x$ in this
graphical language. The flux variables $l_{x,\nu}$ and $\overline{l}_{x,\nu}$
are assigned to the links of the lattice. In Fig.~\ref{variablesgrafical} we
show a site $x$, as well as its two neighbor sites along the 
$\nu$ direction, and indicate the flux variables assigned to the corresponding
links. The monomer variables $s_{x}$ and $\overline{s}_{x}$ are located at the
sites, and again Fig.~\ref{variablesgrafical} illustrates the assignment. 

In our graphical language we use arrows and triangles to represent the flux and
monomers (see Fig.~\ref{fluxexamples} below). For a flux variable  $l_{x,\nu} = n$ we
put $n$ arrows pointing in positive $\nu$ direction on the link between $x$ and $x +
\hat{\nu}$. For $\overline{l}_{x,\nu} = n$ we use  $n$ arrows pointing in negative
$\nu$-direction, since the $\overline{l}_{x,\nu}$ enter with a relative minus sign in
the expression  (\ref{localflux}) for the local flux $f_x$. Similarly, for a monomer
value $s_{x} = n$ we use $n$ outwards (i.e., away from $x$) 
pointing triangles, while the $\overline{s}_{x}$
are represented by inwards pointing triangles. We stress that
Fig.~\ref{fluxexamples} shows only flux-monomer configurations for a two-dimensional
sub-lattice, the $1$-$2$ plane, and the vertical dimension in the plot is used to
display the triangle symbols for the $s_x$ (pointing outwards) and $\overline{s}_{x}$ 
(pointing inwards) monomers.

\begin{figure}[t]
\begin{center}
\includegraphics[height=25mm,clip]{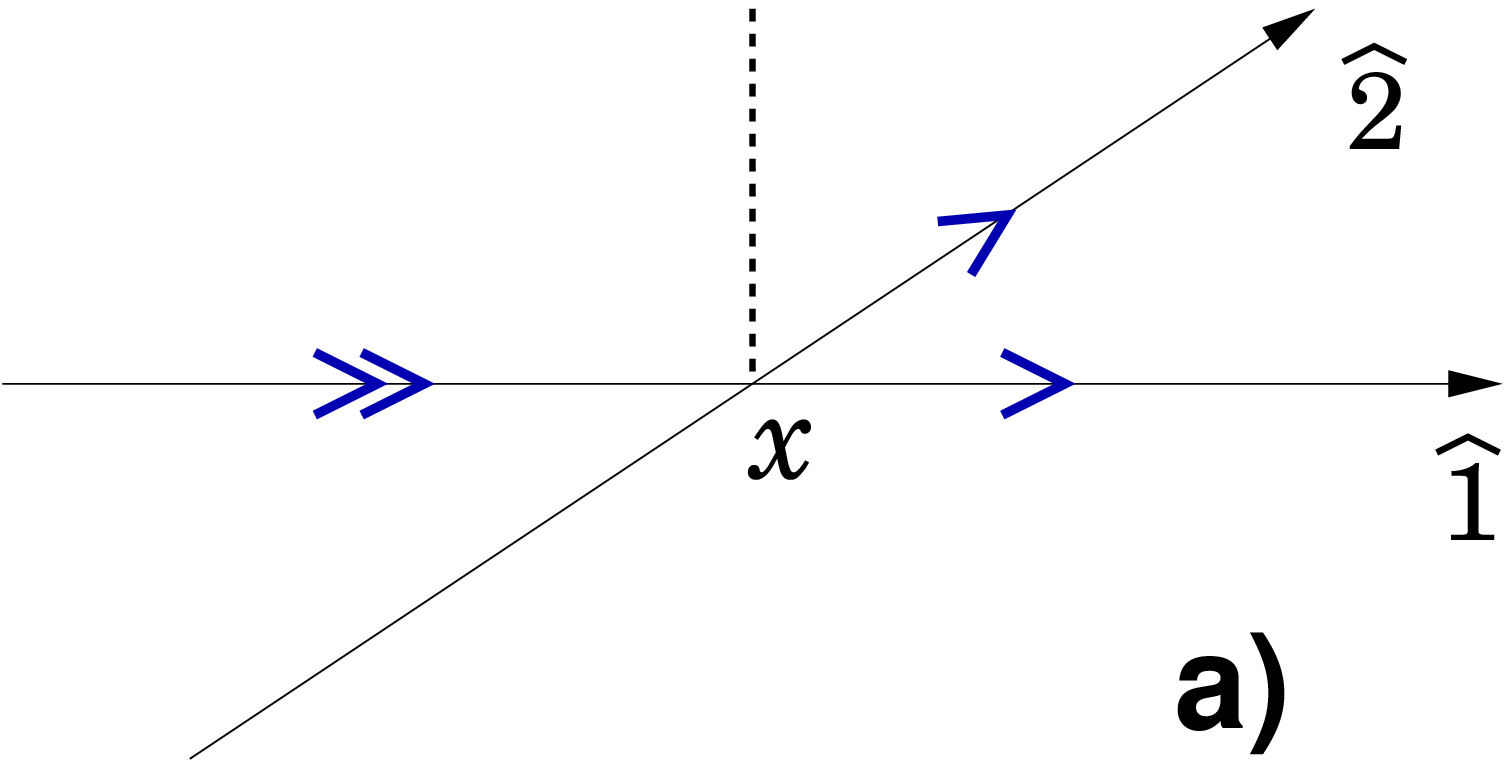}
\hskip10mm
\includegraphics[height=25mm,clip]{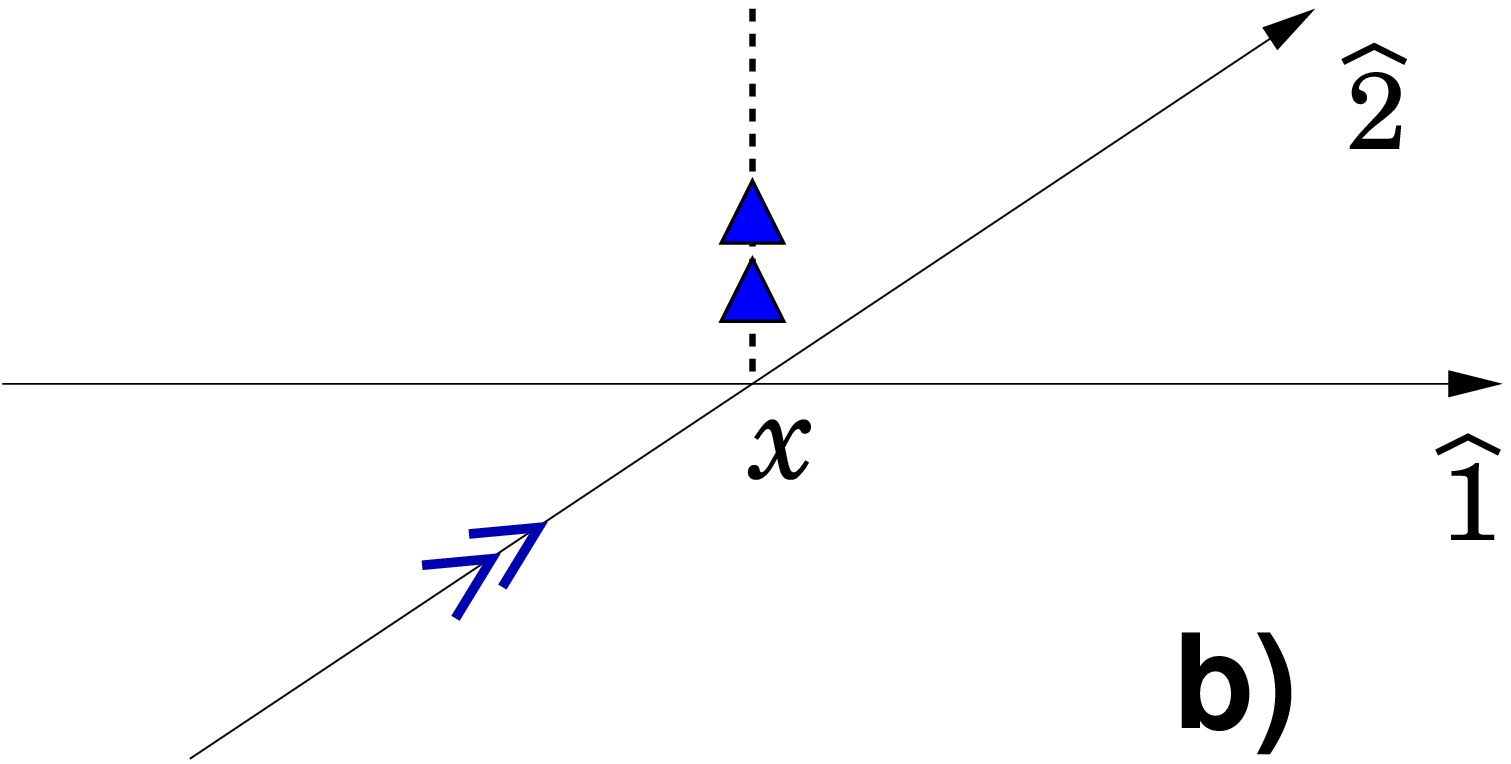}
\vskip15mm
\includegraphics[height=25mm,clip]{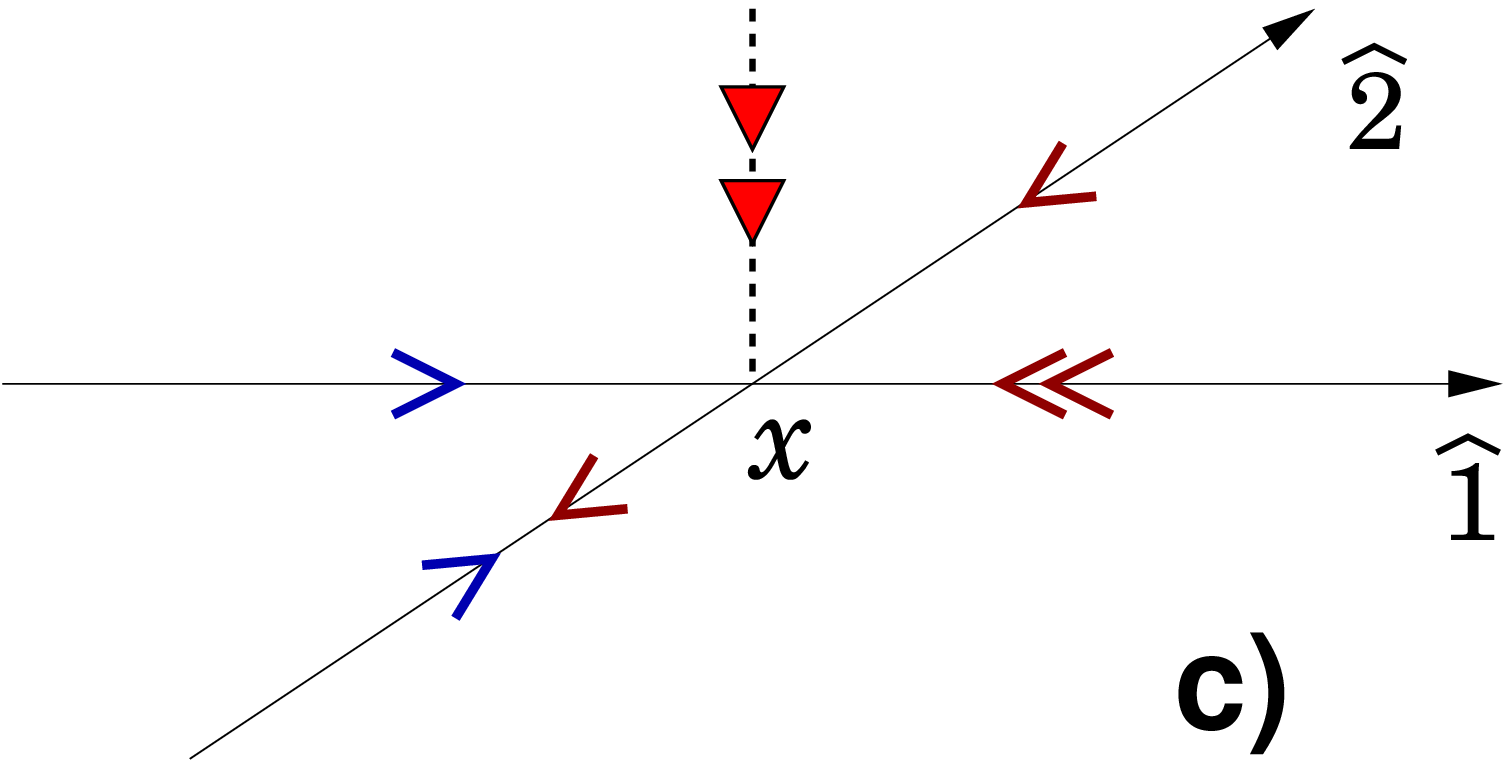}
\hskip10mm
\includegraphics[height=25mm,clip]{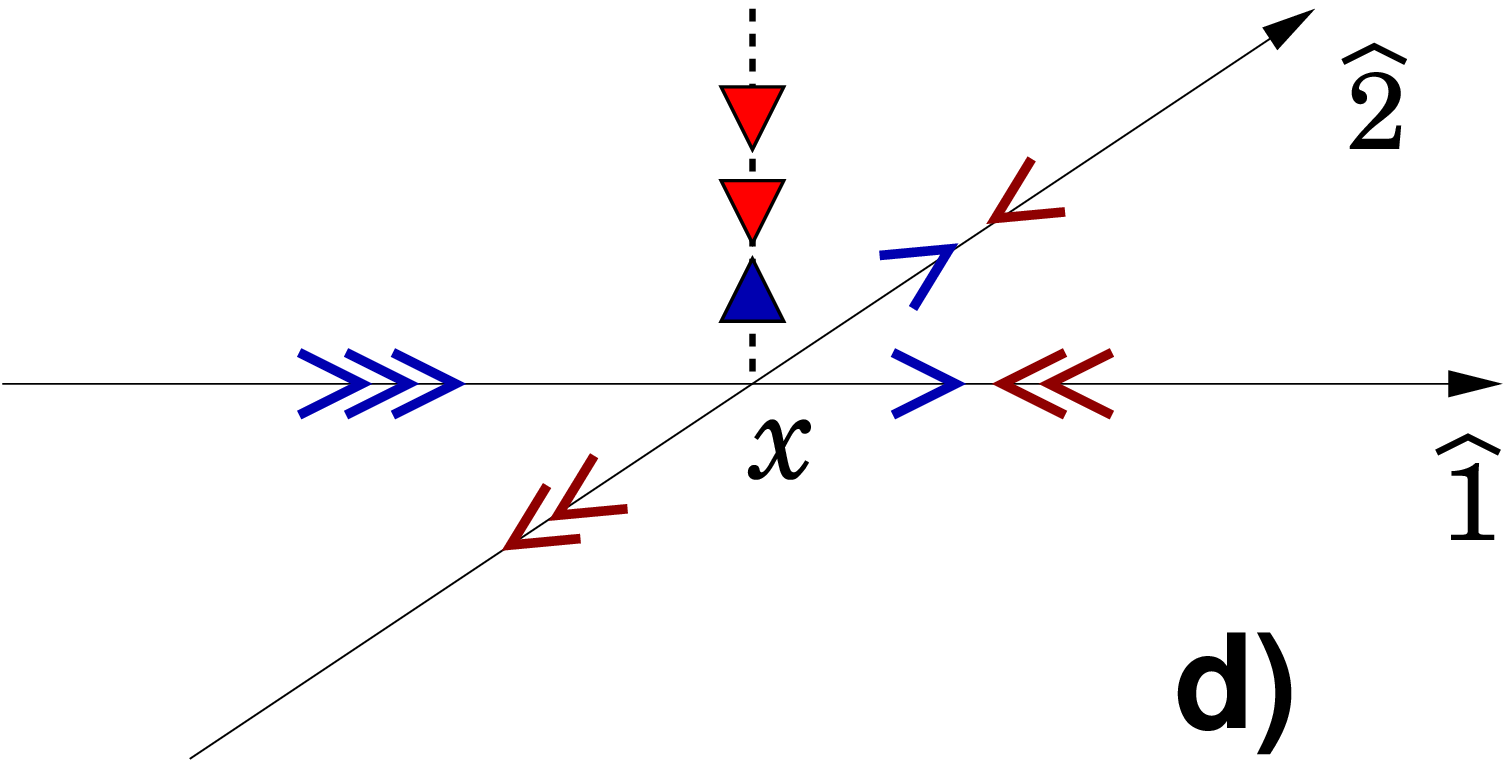}
\end{center}
\caption{Examples of admissible flux-monomer configurations at a single site
$x$. For clarity we show only configurations in the $1$-$2$ plane 
(see the axis labels). A flux variable of $l_{x,\nu} = n$ is represented by 
$n$ arrows pointing in positive $\nu$-direction, while the 
$\overline{l}_{x,\nu}$ point in negative $\nu$-direction. A monomer variable 
$s_{x} = n$ is represented by $n$ outwards pointing (i.e., pointing away from the site $x$)
triangles, while for the 
$\overline{s}_{x}$ we use inwards pointing triangles. The details of the four
examples a) -- d) are discussed in the text.  
\label{fluxexamples}}
\end{figure}

In Fig.~\ref{fluxexamples} we show four examples of admissible flux-monomer
net fluxes $f_x$ at a site $x$. The first example (Fig.~\ref{fluxexamples}a)
is a double line of flux entering from the negative 1-direction, which then
splits at $x$ and two single fluxes exit in 1- and in 2-direction. The 
non-vanishing variables are $l_{x-\hat{1},1} = 2, l_{x,1} = 1, l_{x,2} = 1$.
All other flux and monomer variables attached to $x$ vanish. The total net flux
is $f_x = 0$. In Fig.~\ref{fluxexamples}b a double line of $l$-flux enters and 
is compensated by two units of the monomer variable ($l_{x-\hat{2},2} = 2, s_x = 2$, $f_x
= 0$). Fig.~\ref{fluxexamples}c shows an example of a non-zero total flux $f_x
= -6$ generated by a combination of nonvanishing $l$- and $\overline{l}$-fluxes,
as well as monomers ($l_{x-\hat{1},1} = 1, l_{x-\hat{2},2} = 1,
\overline{l}_{x-\hat{2},2} = 1, \overline{l}_{x,1} = 2, 
\overline{l}_{x,2} = 1, \overline{s}_x = 2$). 
Finally Fig.~\ref{fluxexamples}d
is an example with $f_x = -3$, described by $l_{x-\hat{1},1} = 3,   
\overline{l}_{x-\hat{2},2} = 2, l_{x,1} = 1, l_{x,2} = 1, 
\overline{l}_{x,1} = 2, 
\overline{l}_{x,2} = 1, s_x = 1, \overline{s}_x = 2$.  The
configurations in the sum $\sum_{\{l,\overline{l},s,\overline{s}\}}$ are then
obtained by combining admissible flux arrangements at all sites $x$.

The graphical representation will be useful for developing generalized 
Prokof'ev-Svistunov worm algorithms \cite{worm} which may be used to update
the flux representation efficiently \cite{wip}. Furthermore, one can expand 
the partition sum (\ref{zfinal}) for small $\tau$. Such a result will be
important to check the outcome of numerical simulations (see also
\cite{DelEveGat}), and the graphical representation is a highly welcome tool for
organizing the terms of such an expansion.

\vskip8mm
\noindent
{\Large Flux representation of observables}
\vskip3mm
\noindent
For a successful application of the flux representation of the effective theory
in a Monte Carlo simulation also the observables have to be expressed in terms
of flux and monomer variables. We here briefly discuss this issue for simple
bulk and fluctuation observables. In particular we focus on the expectation value
$\langle P \rangle$ of the Polyakov loop, where $P = \sum_x \mathrm{Tr} \, P(x)$, 
the corresponding susceptibility 
$\chi_P = \langle P^2 \rangle - \langle P \rangle^2$,
the internal energy $U = \langle S \rangle$ (where $S$ is the action
(\ref{action_original})), and the heat capacity 
$C = \langle S^2 \rangle - U^2$. These quantities may be obtained as 
simple derivatives of the partition sum,
\begin{eqnarray}
\hspace*{-6mm}&&\langle P \rangle \; = \; \frac{1}{Z} \frac{\partial}{\partial \eta} \, Z
\; , 
\label{simpleobs} \\ 
\hspace*{-6mm}&&\chi_P \; = \; \frac{1}{Z} \frac{\partial^2}{\partial \eta^2} \, Z \, - \,
\langle P \rangle^2
\; , 
\nonumber 
\\
\hspace*{-6mm}&& U \, = \,  \frac{1}{Z} \, \bigg[ \tau \frac{\partial}{\partial \tau} \, + \, 
\eta \frac{\partial}{\partial \eta} \, + \, 
\overline{\eta} \frac{ \partial}{\partial \overline{\eta}} \bigg] \, Z \; ,
\nonumber
\\
\hspace*{-6mm}&& C  =   \frac{1}{Z} \bigg[ 
\tau^2 \frac{\partial^2}{\partial \tau^2} +  
\eta^2 \frac{\partial}{\partial \eta^2} + 
\overline{\eta}^2 \frac{\partial}{\partial \overline{\eta}^2}  + 
2\tau\eta \frac{\partial^2}{\partial \tau \partial \eta} +  
2\tau \overline{\eta} \frac{\partial^2}{\partial \tau \partial \overline{\eta}} + 
2\eta \overline{\eta} \frac{\partial^2}{\partial \eta \partial \overline{\eta}} 
\bigg] Z  - U^2 .
\nonumber
\end{eqnarray}
From the final form (\ref{zfinal}) of the partition sum it is obvious that the
derivatives in (\ref{simpleobs}) may be pulled under the sum over all flux and
monomer configurations and there only act on the weight factor 
${\cal W}[l,\overline{l},s,\overline{s}]$, since the constraint is independent
of $\tau, \eta$ and $\overline{\eta}$. To perform the needed
derivatives we write the weight factor as
\begin{equation}
{\cal W}[l,\overline{l},s,\overline{s}] \; = \; 
\tau^{L + \overline{L}} \; \eta^{S}
\; \overline{\eta}^{\overline{S}} \;
\left(
\prod_{x,\nu} \; 
\frac{1}{l_{x,\nu}! \; \overline{l}_{x,\nu}!}
\right) \left( \prod_{x} \;
\frac{1}{s_x! \; \overline{s}_x!}  
\right) \, {\cal I}[l,\overline{l},s,\overline{s}] \; ,
\label{weight2}
\end{equation}
where we introduced abbreviations for the sums of flux and monomer variables
\begin{equation}
L \; = \; \sum_{x,\nu} l_{x,\nu} \; , \; \; \; 
\overline{L} \; = \; \sum_{x,\nu} \overline{l}_{x,\nu} \; , \; \; \;
S \; = \; \sum_{x} s_{x} \; , \; \; \;
\overline{S} \; = \; \sum_{x} \overline{s}_{x} \; .
\label{sums}
\end{equation}
Using the weight factor (\ref{weight2}) one immediately obtaines the
following relations:
\begin{equation}
\frac{\partial}{\partial \eta} {\cal W}[l,\overline{l},s,\overline{s}] \; = \;
\frac{S}{\eta} \, {\cal W}[l,\overline{l},s,\overline{s}] \; , \; \; \;
\frac{\partial^2}{\partial \eta^2} {\cal W}[l,\overline{l},s,\overline{s}] \; = \;
\frac{S^2-S}{\eta^2} \, {\cal W}[l,\overline{l},s,\overline{s}] \; ,
\label{derivatives}
\end{equation}
for the derivatives with respect to $\eta$, and similar relations
for the derivatives with respect to the other parameters. The important 
aspect of the expressions (\ref{derivatives}) is that the derivatives of the
weight factor may be written as the unchanged weight 
${\cal W}[l,\overline{l},s,\overline{s}]$, multiplied with
factors built from the parameters $\eta, \overline{\eta}, \tau$ and the
sums (\ref{sums}) of flux- and monomer variables. Reinserting these relations we
obtain rather simple expressions for our observables (\ref{simpleobs}),
\begin{eqnarray}
&& \langle P \rangle \; = \; \frac{1}{\eta} \langle S \rangle \; ,
\label{bulkobsfinal} \\
&& \chi_P \; = \; \frac{1}{\eta^2} \, 
\Big[ \, \langle S^2 - S \rangle \; - \; \langle S \rangle^2 \, \Big] \; ,
\nonumber \\
&& U \; = \;
\langle \, L + \overline{L} + S + \overline{S} \, \rangle \; ,
\nonumber \\
&& C \; = \;
\langle \, ( \, L + \overline{L} + S + \overline{S} \,)^2 \, \rangle \, - \, U \, - \, U^2 \; .
\nonumber
\end{eqnarray}
Obviously all the bulk and fluctuation observables we list here can be expressed in terms of
the (summed) flux and monomer occupation numbers from Eq.~(\ref{sums}), as well as
the corresponding fluctuations. 

As a matter of fact, in a similar way also correlators can be expressed in terms of
the flux and monomer variables. We briefly discuss this for the example of the
Polyakov loop correlator $\langle \, \mathrm{Tr} P(x) \,  \mathrm{Tr} P(y)^\dagger
\, \rangle$. The two factors $\mathrm{Tr} P(x)$ and $\mathrm{Tr} P(y)^\dagger$ can
be generated from the partition sum by using parameters $\eta(x)$ and
$\overline{\eta}(x)$ in (\ref{action_original}), which are independent parameters 
for all sites $x$ of the lattice. The mapping of this generalized model with locally
varying parameters to the
flux-monomer representation goes through unchanged. The local 
parameters $\eta(x)$ and
$\overline{\eta}(x)$ can now be used as sources and the correlator 
$\langle \, \mathrm{Tr} P(x) \,  \mathrm{Tr} P(y)^\dagger \, \rangle$
is obtained via two partial derivatives. In the end one sets all 
parameters to the original values, i.e., $\eta(x) = \eta$ and
$\overline{\eta}(x) = \overline{\eta}$ for all $x$. One ends up with
\begin{equation}
\langle \, \mathrm{Tr} P(x) \,  \mathrm{Tr} P(y)^\dagger \, \rangle \; = \;
\frac{1}{Z}\, \frac{\partial^2}{\partial \eta(x) \, \partial \overline{\eta}(y)}
\, Z \; \Bigg|_{{\eta(x) = \eta \atop  \overline{\eta}(x) = \overline{\eta}}} \; = \;
\; \frac{1}{\eta \, \overline{\eta}} \, \langle \, s_x \, \overline{s}_y  \, \rangle \; .
\end{equation}
We find that the Polyakov loop correlator turns into a correlator of the
monomer numbers at the corresponding sites. In a simulation of the flux-monomer representation
with a worm algorithm one can evaluate this correlator by directly sampling the
starting point of the worm correlated with its actual position and with this
improved estimator one expects to get an excellent signal. 

\vskip8mm
\noindent
{\Large Summary, discussion, outlook}
\vskip3mm
\noindent
In this article we study an effective Polyakov loop theory for QCD thermodynamics. 
The model may be obtained from a strong coupling expansion, combined with a hopping
expansion for the fermion determinant. It contains the leading center symmetric 
and center symmetry breaking terms. At non-zero chemical potential the model inherits
the complex phase problem of finite density QCD. The complex phase problem prohibits
a direct Monte Carlo simulation in the standard representation.

Using high temperature expansion techniques, we map the partition sum of the
effective Polyakov loop model to a representation where the dynamical degrees of
freedom are integer valued flux variables attached to the links of the lattice, and
integer valued monomer variables at the sites. The flux-monomer configurations are subject
to a constraint which forces the local net flux at each site to have vanishing triality,
i.e., the flux has to be a multiple of three. All admissible configurations come with
a real and positive weight factor, and thus in the flux-monomer representation the
complex phase problem is solved.

For systems with such a flux representation generalizations of the                                        
Prokof'ev-Svistunov worm algorithm \cite{worm} allow for a numerical Monte Carlo
simulation of the model. As a matter of fact, for a similar QCD-related system a
generalized worm algorithm was used to effectively explore the full temperature
and chemical potential parameter space, also in regions where the complex phase problem
is severe \cite{DelEveGat}. We expect \cite{wip} that for the system discussed here, a worm
algorithm with similar efficiency can be implemented on the basis of the
new flux-monomer representation given here. Again it should be possible to
explore the full temperature-density phase diagram. Such an analysis will provide
interesting insight into various aspects of the phase structure of QCD-related
systems.

Besides the improvement of our physical understanding, a simulation based on the
new flux representation will also have important technical applications. The 
models can serve as a prototype system, which other
approaches to QCD with non-zero chemical potential can be tested against. Various
expansions in $\mu$, imaginary chemical potential techniques, as well as results
from reweighting (see \cite{finchemreview} for reviews) could be compared to the outcome of a Monte Carlo simulation in
the flux representation, where the simulation is not tainted by uncontrolled
effects from a complex phase problem. 

However, not only lattice techniques should be compared to the prototype system
presented here. It is straightforward to write down a continuum counterpart of
the action (\ref{action_original}), and the system could also be analyzed with various
continuum techniques, in particular functional methods \cite{frg}. A comparison of these
results with the lattice would provide important insights on the assumptions and
the reliability of the continuum techniques.

\vskip5mm
\noindent
{\Large Acknowledgments}
\vskip4mm
\noindent
The author thanks Gert Aarts, Ydalia Delgado-Mercado, Hans Gerd Evertz, Christian Lang,
Bernd-Jochen Schaefer, Owe Philipsen and
Andreas Wipf for stimulating discussions and remarks on the literature.

\end{document}